\documentclass[preprint]{revtex4}   
\usepackage{amsfonts}
\usepackage{amsmath}
\usepackage{cancel}
\usepackage{graphicx}
\usepackage{mathcomp}
\usepackage{color}
\usepackage{ulem}
\usepackage{mathtools}

\usepackage[utf8]{inputenc}

\allowdisplaybreaks

\newif\iffigures
\figurestrue

\renewcommand{\vec}{\mathbf}

\def\undertilde#1{\mathord{\vtop{\ialign{##\crcr
$\hfil\displaystyle{#1}\hfil$\crcr\noalign{\kern1.5pt\nointerlineskip}
$\hfil\widetilde{}\hfil$\crcr\noalign{\kern1.5pt}}}}}

\begin{document}
\title{Theoretical investigation of a miniature microwave driven plasma jet}

\author{Michael Klute$^{1}$}
\author{Horia-Eugen Porteanu$^{2}$}
\author{Ilija Stefanovi\'{c}$^{3}$}
\author{Wolfgang Heinrich$^{2}$}
\author{Peter Awakowicz$^{3}$}
\author{Ralf Peter Brinkmann$^{1}$}

\affiliation{$^{1}$Institute for Theoretical Electrical Engineering, Ruhr University Bochum, D-44780 Bochum, Germany}
\affiliation{$^{2}$Microwave Department, Ferdinand-Braun-Institut, Berlin Germany}
\affiliation{$^{3}$Electrical Engineering and Plasma Technology,
Ruhr University Bochum, Germany}
\date{\today}

\begin{abstract}
Microwave and radio frequency driven plasmas jets play an important role in many technical applications. They are usually operated in a capacitive mode known as E-mode. As a new plasma source the MMWICP (Miniature Micro Wave ICP) has been proposed, a small scale plasma jet with inductive coupling based on a specially designed resonator that acts as an LC-resonance circuit. This work presents a theoretical model of the new device, based on a series representation of the electromagnetic field in the resonator and the volume integrated (global) model for the loss processes within the plasma. An infinite number of modes can be found ordered by the azimuthal wave number m. These modes essentially determine the electromagnetic behavior of the system and differ from ordinary cavity modes. The mode $m=0$  can be identified with the inductive mode and is called H-mode, the mode $m=1$ is the capacitive mode and is called E-mode. Both modes refer to different operating regimes, which are separated by different values of the plasma parameters. In a second step the matching network and its characteristics are taken into account in order to find stable equilibrium points and possible hysteresis effects. As main result, the feasibility of inductive power coupling for the MMWICP resonator is shown.
\end{abstract}

\maketitle

\pagebreak


\section{Introduction}

\begin{figure}[htbp]
\centering
\iffigures		
   		\includegraphics[width=0.99\textwidth]{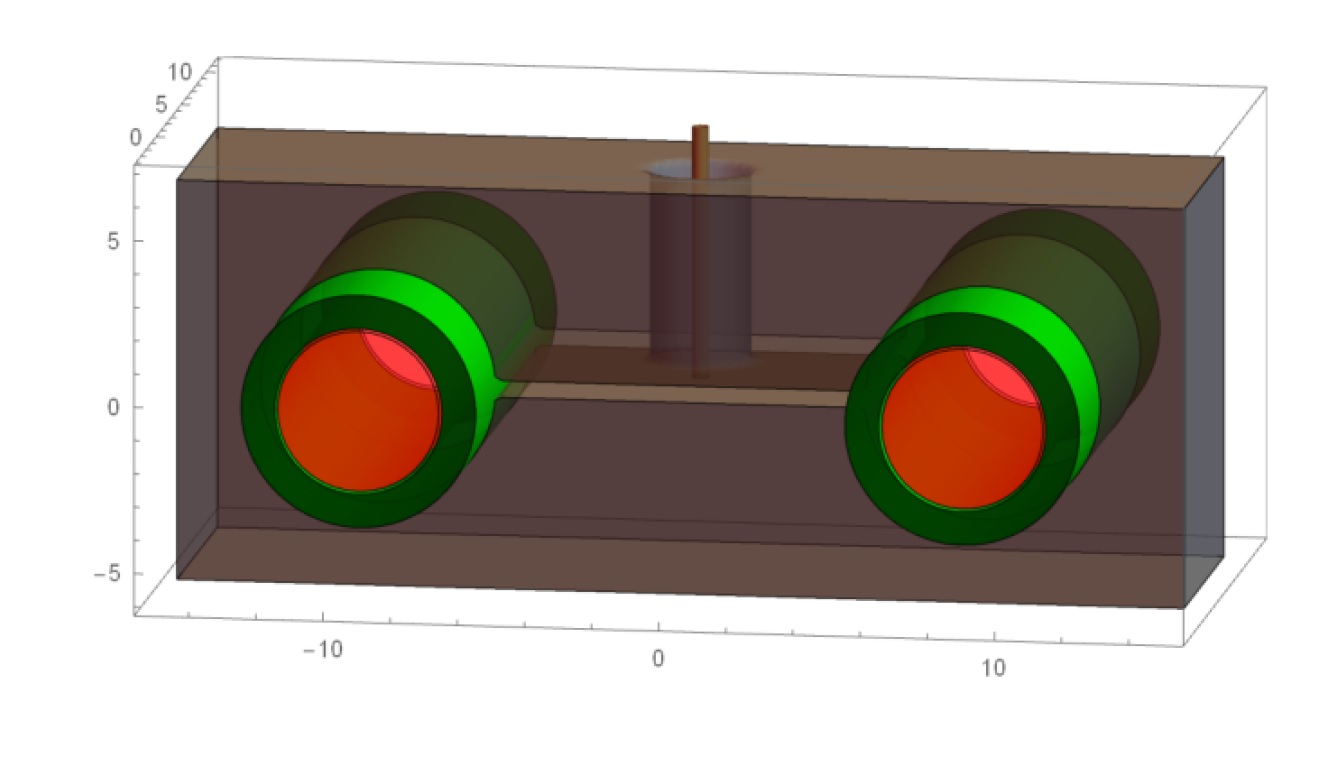}
\else 
		
\fi 
\caption{Visualization of the MMWICP plasma jet. The green color represents the dielectric, while the red color represents the plasma itself.}
	\label{fig:paperfig1}
\end{figure}

Plasma processes can be classified into ``direct'' and ``remote''. In direct plasma processes, the generation and application of the plasma are co-located: A work-piece is introduced into a plasma chamber and processed therein. 
In remote plasma processes, plasma generation and plasma application are spatially separated. 
In the source, the plasma is only generated; \linebreak 
from there it emerges as a beam of energetic particles and strikes the object that is being treated. Remote plasma processes have significant advantages: The source and the remote plasma are almost independent in their properties and can be optimized separately. The sources themselves are often quite small (``miniaturized''), and can be handled very flexibly. They can be designed for a wide pressure range: 
low pressure (about $10$-$10^3\,\mathrm{Pa}$) is of interest, e.g.,  for plasma enhanced thin film deposition, \linebreak
atmospheric pressure ($10^5\,\mathrm{Pa}$) allows medical and environmental applications. Among the sources suitable for remote processes, radio frequency (RF) or microwave (MW) operated plasma jets play a special role. They are usually operated in a capacitive, or E-Mode: the RF or MW power is applied to the electrodes, resulting in a strong electric field in the interior of the jet. If the field strength exceeds the breakdown value, the discharge ignites and a plasma forms. It comes to the formation of a quasi neutral bulk and a strongly electron depleted boundary sheath. Capacitive coupling allows plasma jets with a simple structure and favorable ignition behavior. However, it also has considerable disadvantages: in the operating state, only the bulk voltage contributes to the electron heating; this limits the efficiency and thus also the achievable plasma density. On the other hand, the voltage across the sheath increases the kinetic energy of the ions, which are striking the electrodes. The high ion energy also causes a strong erosion and heating of the electrodes, up to the ignition of a parasitic arc. These disadvantages motivate the search for an alternative form of coupling with focus on inductive coupling. Instead of using electrodes, the typical design of an inductively coupled plasma (ICP) reactor consists of either a planar or helical coil, which is located outside the plasma chamber and driven by a RF or MW signal. A time varying magnetic field is created by the coil. The magnetic field in turn induces a strong current within the plasma, which is gaining energy in this way. The inductive coupling is also known as H-Mode. ICPs show much brighter light emissions, up to 10-times higher electron densities and a lower plasma potential \cite{ref1}. In the same time they show lower ion energies, resulting in less erosive effects, such as ion sputtering \cite{ref2},\cite{ref3}. Successful examples for small scale systems are RF induction lamps \cite{ref4}, \cite{ref5}. Attempts have been made to realize inductive coupling also for plasma jets. A successful example for a large scale ICP jet is the $\textit{induction-coupled plasma torch}$ \cite{ref6}. First attempts to realize small scale ICP jets have shown to be only partially successful; doubts were expressed by the authors if the H-mode could be achieved or not \cite{ref7}-\cite{ref9}. It is interesting to look at the following expression, taken from the textbook by Lieberman and Lichtenberg \cite{ref10}: The power, $P_{RF}$, coupled to an MW driven ICP with chamber radius $R_{1}$ and a balanced, cylindrical geometry (radius $R_{1}$ $\approx$ height $H$) can be expressed as a function of skindepth $\lambda_s$, inductor current $I$, effective scattering frequency $\nu$, permeability of free space $\mu_0$ and winding number $N$:
\begin{align}
P_{RF}= \begin{cases}
     \mu_0 \nu N^2 I^2 R_{1}^3/\lambda_s^2 & \text{for } \lambda_s \gtrsim R_{1} \\
     \mu_0 \nu N^2 I^2 \lambda_s & \text{for } \lambda_s  \lesssim R_{1}.
   \end{cases}
\end{align}
Apparently a small scale ICP can only be operated efficiently if the chamber radius $R$ and the skindepth $\lambda_s$ and are at least of comparable size (or better $\lambda_s < R$ ). Thus, high densities are required as $\lambda_s$ is proportional to $n_{e}^{-1/2}$. Here $n_e$ denotes the plasma electron density. Often the winding-number $N$ was increased to improve the coupling efficiency. This leads to a higher voltage drop at the coil and a decreased resonance frequency, limiting the parameter range of the discharge. 
In a previous work of Porteanu et al. \cite{ref1} a different approach was developed. A LC-resonance circuit was implemented into a resonator. The design is depicted in Figs \ref{fig:paperfig1} and \ref{fig:bild3}. The typical voltage drop (due to the coil) is strongly limited for this design. Thus, the occurrence of an undesired capacitive discharge can be suppressed. The system losses are minimized and an optimum of the energy transfer to the plasma is achieved. Some experimental results have been gathered already from a prototype model \cite{ref1}, \cite{ref11}, \cite{ref12}.
\newline
\newline
The purpose of this paper is to theoretically examine the prototype with focus on its electromagnetic behavior. In chapter 2 a detailed description of the reactor prototype and its electrical matching network is given. In chapter 3 an analytical description of the electromagnetic fields is given based on a series representation. The results differ from ordinary cavity modes. A linear model for the power losses is added to consider the power balance of the system. In chapter 4 fundamental insights into the system is presented regarding stable operating points, operating regimes, hysteresis effects and the pattern of field lines. The terms E- and H-mode will be applied to the system. The general feasibility of the inductive coupling for a small scale plasma jet is shown. A comparison to existing experiments is limited to the electromagnetic aspects as experimental results are only available for nitrogen yet. The paper concludes with chapter 5, where the results are summarized, discussed and an outlook to further modifications is given. 
\pagebreak

\color{black}

\section{The demonstrator and its idealization}
\begin{figure}
\centering
\iffigures		
   	\includegraphics[width=0.99 \textwidth]{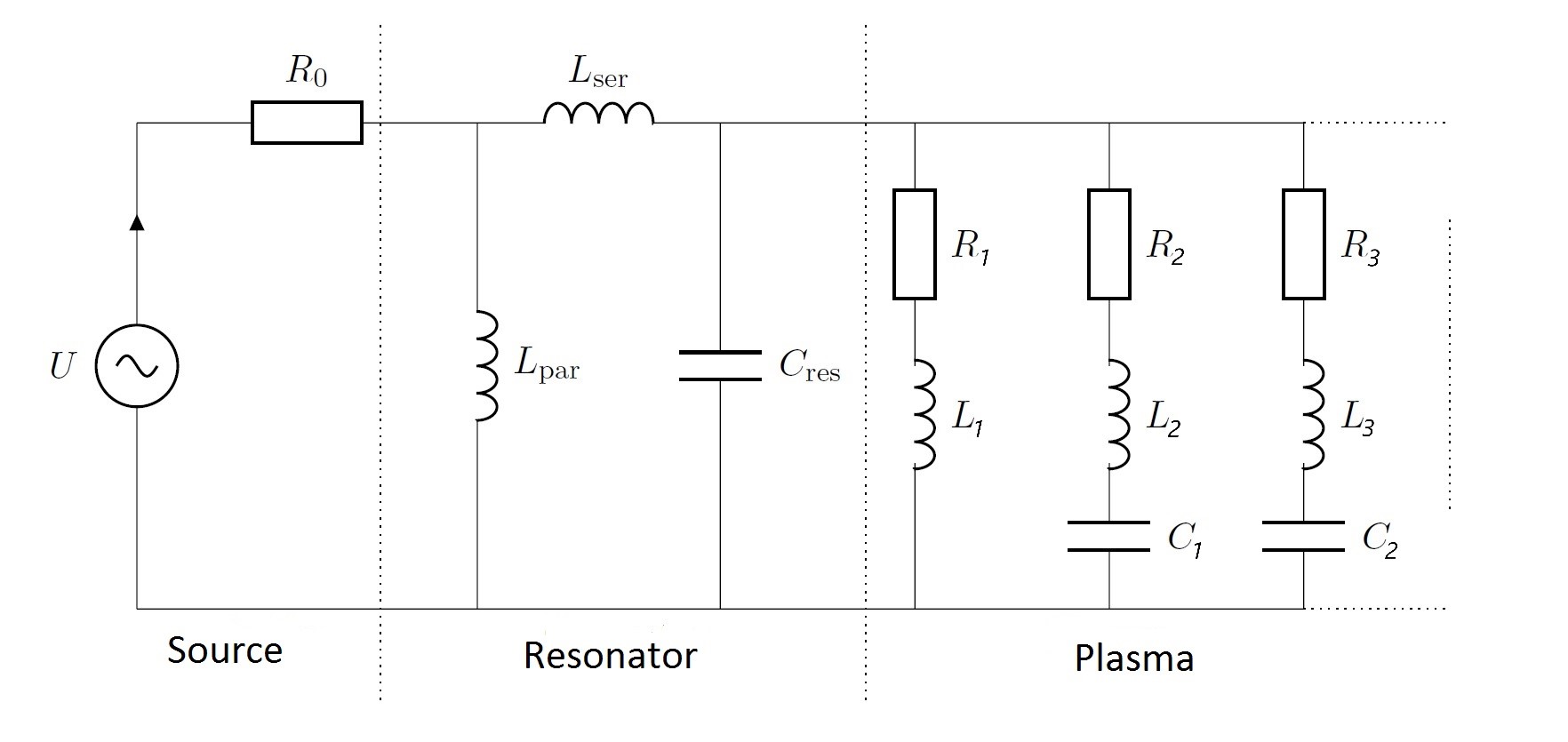}
\else 
		
\fi 
		
	\caption{Lumped element representation of the system, consisting of the power source, resonator and the plasma itself. The matching network is included in the resonator. The capacitive coupling is taken into account as LCR-circuits (for $m=1,2,...$) while the inductive coupling is represented as an RL-circuit ($m=0$).}
	\label{fig:Fig1}
\end{figure}

\begin{figure}
	\centering
\iffigures		
   		\includegraphics[width=0.90 \textwidth]{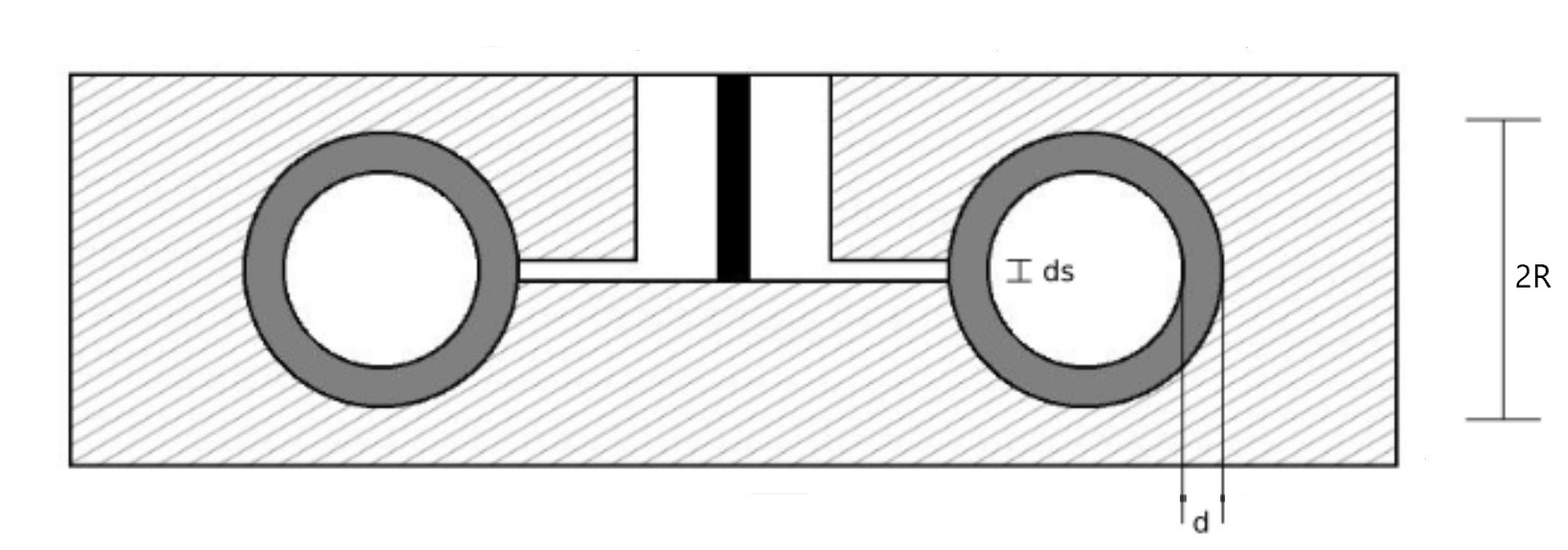}
\else 
		
\fi 
		
	\caption{A schematic cross section of the reactor, where $R$ represents the radius of the resonator chamber, $d$ the wall thickness of the dielectric tube and $d_S$ the gap distance.}
	\label{fig:bild3}
\end{figure}

A schematic drawing of the plasma reactor is shown in Fig \ref{fig:bild3}. The design is based on a resonator made of a solid block of copper with a length of $\textrm{l}_1=34\,\textrm{mm}$, a height of $\textrm{h}=10\,\textrm{mm}$ and a depth of $\textrm{l}_2=8\,\textrm{mm}$ \cite{ref1}. There are two parallel cavities with radius $\textrm{R}=3.5\,\textrm{mm}$ in the copper block. Both are connected by a gap of length $\textrm{l}_s=11\,\textrm{mm}$ and width $\textrm{d}_s=0.22\,\textrm{mm}$. Two dielectric tubes with wall thickness $\textrm{d}=1\,\textrm{mm}$, through which a gas flows, are guided through the parallel cavities. The plasma is ignited within the tubes. The resonator is shielded with a solid $8\,\textrm{mm}$ thick housing of aluminum. The resonator is driven by an external MW amplifier with frequency $f= 2.45\times 10^{9}\,\textrm{Hz}$ (angular frequency $\omega= 1.54\times 10^{10}\,\textrm{s}^{-1}$) and an amplitude expressed as output voltage $u_\textrm{S}$). Due to the ratio between the the dimensions ($2\times \textrm{R}=7\,\textrm{mm}$) and the wavelength of the microwave signal ($\lambda_\textrm{RF}\approx 12\,\textrm{cm}$) cavity modes are not excited in the resonator.\newline
\newline
For the proposed theoretical description the following idealizations is made: An electromagnetic regime will be assumed within the resonator cavities and an electrostatic regime will be assumed within the gap connecting the resonators cavities. In particular, the system of the two cavities and the gap in between is considered as a system of coupled waveguides; electromagnetic effects can not propagate into the gap. In conclusion a simplified description using lumped elements is possible: Both cavities are regarded as coils with one winding and with the individual inductance $L_R=6\,\textrm{nH}$, while the gap is regarded as a capacitor with capacity $C_{\textrm{res}}=3\,\rm{pF}$. These elements form an idealized LC-circuit with resonance frequency $f_\textrm{Res}\approx 2.2\times 10^{9}\textrm{Hz}$. All interactions between the two cavities and the gap are limited to the lumped element model. \newline
\newline
During operation the jet works as a "`double jet"' with two parallel discharges and two separate plasma effluents. As the gap area is limited to an electrostatic regime, no electromagnetic interaction between the two parallel discharges are taken into account. Therefore the full electromagnetic calculation is performed for one discharge only ("`single jet"' instead of "`double jet"'). In addition, inside the cavity three zones are defined with respect to the radius r. In the range $r=0\ldots (R-\delta-d)$ it is the plasma bulk, $r=(R-\delta-d)\ldots (R-d)$ represents the plasma sheath zone and $r=(R-d)\ldots R$ represents the dielectric zone. Furthermore, the electromagnetic field is assumed to be invariant along the axis of the tube/gasflow. This follows from the aspect ratio of the resonator ($R<\textrm{l}_2$) in general, as well as from the ratio between $\textrm{l}_2$ and the skin depth in particular ($\lambda_s\ll\textrm{l}_2$).\newline
\newline
An expression for the complex admittance of one single discharge $Y_\textrm{P}(n_{\textrm{e}},\omega)$ is derived from calculations, so the plasma can be incorporated into the lumped element model. In order to efficiently connect the resonator to a microwave source (with the inner impedance of $\textrm{Z}_{0}=50\,\Omega$), an impedance transformation is required. Such a transformation is realized by the external inductances $L_\textrm{ser}$ and $L_\textrm{par}$ and the resonators capacitor $C_\textrm{res}$, which are forming a matching network. The resulting admittance $Y_\textrm{S}(\omega,n_{\textrm{e}})$ of a "`single jet"'-system follows from the circuit in the experiment and is given by: 
\begin{align}\label{eq:impedance}
Y_\textrm{S}(\omega,n_{\textrm{e}})=\left(\frac{1}{(Y_\textrm{P}(n_{\textrm{ne}},\omega)+i\omega C_{\textrm{res}})^{-1}+i\omega L_{\textrm{ser}}} + \frac{1}{i\omega L_{\textrm{par}}} \right). 
\end{align}
The resulting circuit diagram of the system is completed by the lumped element representation of the plasma itself (one inductive branch and an infinite number of capacitive branches) and shown in Fig. \ref{fig:Fig1}. The considered pressure is $\textrm{p}=100\,\textrm{Pa}$.

\pagebreak

\section{Electromagnetic model (of a single Jet)}
\subsection{Ansatz}
Cylindrical coordinates, $(r,\phi,z)$, are used for the calculation, where $r$, $\phi$ and $z$ have their usual meanings. The $z$-axis is centered in the middle of the cavity and the gap will be located at $\phi=0$. The dimensions of the resonator only allow the excitation of transversal-electric modes (TE-Mode) with a magnetic field pointing in $z$-direction. Due to the gap there is no azimuthal symmetry and the transversal fields depend on $r$ as well as on $\phi$. Consequently the magnetic field $\vec{B}$ and the electric field $\vec{E}$ can be written as follows, where $n$ is a zone index that takes the values p (plasma), s (sheath) and d (dielectric):
\begin{align}
\vec{B}^{(n)}&=\textrm{Re}(\underline{B}_z^{(n)}(r,\phi)\,\textrm{e}^{i\omega t}\,\vec{e}_{z}), \\ 
\vec{E}^{(n)}&=\textrm{Re}(\underline{E}_r^{(n)}(r,\phi)\,\textrm{e}^{i\omega t}\,\vec{e}_{r}+\underline{E}_{\phi}^{(n)}(r,\phi)\,\textrm{e}^{i\omega t}\,\vec{e}_{\phi}).
\end{align}
In plasma, additionally, the charge density $\rho$ and the current density $\vec{j}$ have to be considered:
\begin{align}
  \rho  &= \textrm{Re}(\underline{\rho}(r,\phi)\,\textrm{e}^{i\omega t}), \\ 
	\vec{j}&=\textrm{Re}(\underline{j}_r(r,\phi,t)\,\textrm{e}^{i\omega t}\,\vec{e}_{r}+\underline{j}_{\phi}(r,\phi)\,\textrm{e}^{i\omega t}\,\vec{e}_{\phi})
\end{align}
The fields obey the full set of Maxwell equations, with electrical transport specifications for each of the zones.
In the plasma, the equations are as follows, with vacuum permittivity $\epsilon_0$:
\begin{align}
	  &\frac{1}{\mu_0}\frac{1}{r}\frac{\partial \underline{B}_z^{(\mathrm{p})}}{\partial \phi} = \underline{j}_r+\epsilon_0 i \omega \underline{E}_r^{(\mathrm{p})}\label{eq:221}, \\[0.5ex]
	 	&	-\frac{1}{\mu_0}\frac{\partial \underline{B}_z^{(\mathrm{p})}}{\partial r} = \underline{j}_{\phi}+\epsilon_0 i\omega \underline{E}_{\phi}^{(\mathrm{p})},\\[0.5ex]
    & \frac{1}{r}\frac{\partial(r \underline{E}_{\phi}^{(\mathrm{p})}) }{\partial r}-\frac{1}{r}\frac{\partial \underline{E}_r^{(\mathrm{p})}}{\partial \phi} 
	= -i\omega \underline{B}_z^{(\mathrm{p})}, 
\end{align}
\begin{equation}
\begin{split}
\frac{\epsilon_0}{r}\frac{\partial(r \underline{E}_r^{(\mathrm{p})})}{\partial r}+\frac{\epsilon_0}{r}\frac{\partial(\underline{E}_{\phi}^{(\mathrm{p})})}{\partial \phi}&=\underline{\rho}. \\ 
\end{split}
\end{equation}
They are completed by the charge conservation equation,
\begin{align} 
	i\omega\underline{\rho}+\frac{\partial \underline{j}_r}{\partial r}+\frac{1}{r}\frac{\partial \underline{j}_{\phi}}{\partial \phi}=0,
\end{align}
and the equation of motion in the transversal plane, according to the cold plasma model:
\begin{align}
	   &i\omega \underline{j}_r=  \epsilon_0\omega_\mathrm{pe}^2 \underline{E}_r-\nu \underline{j}_r,\\
	   &i\omega \underline{j}_{\phi}=\epsilon_0\omega_\mathrm{pe}^2 \underline{E}_{\phi}-\nu \underline{j}_{\phi}\label{eq:222}.\\
\end{align}
Within the plasma sheath, the Maxwell equations take the following form:
\begin{align}
	  &\frac{1}{\mu_0}\frac{1}{r}\frac{\partial \underline{B}_z^{(\mathrm{s})}}{\partial \phi} =  i \omega \underline{E}_r^{(\mathrm{s})}, \\[0.5ex]
	 	&	-\frac{1}{\mu_0}\frac{\partial \underline{B}_z^{(\mathrm{s})}}{\partial r} =\epsilon_0 i\omega \underline{E}_{\phi}^{(\mathrm{s})},\\[0.5ex]
    & \frac{1}{r}\frac{\partial(r \underline{E}_{\phi}^{(\mathrm{s})}) }{\partial r}-\frac{1}{r}\frac{\partial \underline{E}_r^{(\mathrm{s})}}{\partial \phi} 
	= -i\omega \underline{B}_z^{(\mathrm{s})}. 
\end{align}
\begin{equation}
\begin{split}
\frac{\epsilon_0}{r}\frac{\partial(r \underline{E}_r^{(\mathrm{s})})}{\partial r}+\frac{\epsilon_0}{r}\frac{\partial( \underline{E}_{\phi}^{(\mathrm{s})})}{\partial \phi}&=0. \label{eq2} \\ 
\end{split}
\end{equation}
Within the dielectric, the equations are as follows, with permittivity $\epsilon=\epsilon_0\epsilon_r$:
\begin{align}
	  &\frac{1}{\mu_0}\frac{1}{r}\frac{\partial \underline{B}_z^{(\mathrm{d})}}{\partial \phi} =  i \omega \underline{E}_r^{(\mathrm{d})}, \\[0.5ex]
	 	&	-\frac{1}{\mu_0}\frac{\partial \underline{B}_z^{(\mathrm{d})}}{\partial r} =\epsilon i\omega \underline{E}_{\phi}^{(\mathrm{d})},\\[0.5ex]
    & \frac{1}{r}\frac{\partial(r \underline{E}_{\phi}^{(\mathrm{d})}) }{\partial r}-\frac{1}{r}\frac{\partial \underline{E}_r^{(\mathrm{d})}}{\partial \phi} 
	= -i\omega \underline{B}_z^{(\mathrm{d})}. 
\end{align}
\begin{equation}
\begin{split}
\frac{\epsilon}{r}\frac{\partial(r\underline{E}_r^{(\mathrm{d})})}{\partial r}+\frac{\epsilon}{r}\frac{\partial(\underline{E}_{\phi}^{(\mathrm{d})})}{\partial \phi}&=0. \label{eq2}
\end{split}
\end{equation}

\subsection{General solution}
By using an expansion into a series ansatz, $\underline{B}_z^{(n)}(r,\phi)$ can be rewritten as:
\begin{align}
\underline{B}^{(\mathrm{n})}_z(r,\phi)&=\sum_{m=0}^{\infty}\left(C_m J_m(r\kappa)+D_m N_m(r\kappa)\right)\cos(m\phi).
\end{align}
Here $J_m(r\kappa)$ and $N_m(r\kappa)$ denote the Bessel functions of the first and the second kind and of order $m$ with corresponding constants $C_m$ and $D_m$. The corresponding electric fields $\underline{E}^{(\mathrm{n})}_r(r,\phi)$ and $\underline{E}^{(\mathrm{n})}_\phi(r,\phi)$ are derived from the Maxwell equations. For $\underline{B}_z^{(p)}(r,\phi)$, $\underline{E}_r^{(p)}(r,\phi)$ and $\underline{E}_\phi^{(p)}(r,\phi)$ the constant $D_m$ is zero for all m, as $N_m(r\kappa)$ goes to infinity for $r\rightarrow 0$.

\subsection{Boundary conditions}
The boundary conditions for the electric and magnetic field are considered one by one, beginning with the tangential component of the electric field, where $\vec{n}$ represents the normal vector of the boundary surface. At the boundary between plasma and sheath, as well as between sheath and dielectric, $\underline{E}_{\phi}^{(\mathrm{n})}$ is continuous:
\begin{align}
\underline{E}_{\phi,}^{(\mathrm{p})}(R-d-\delta,\phi)&=\underline{E}_{\phi}^{(\mathrm{s})}(R-d-\delta,\phi)\\
\underline{E}_{\phi}^{(\mathrm{s})}(R-d,\phi)&=\underline{E}_{\phi}^{(\mathrm{d})}(R-d,\phi).
\end{align}
There are no surface currents at the considered boundaries and no magnetization, thus $\vec{B}$ is also continuous:
\begin{align}
\underline{B}^{(\mathrm{p})}_z(R-d-\delta,\phi)&=\underline{B}^{(\mathrm{s})}_z(R-d-\delta,\phi) \\
\underline{B}^{(\mathrm{s})}_z(R-d,\phi)&=\underline{B}^{(\mathrm{d})}_z(R-d,\phi).
\end{align}
For the normal component of the electric field, $\vec{E}\cdot\vec{n}$, the following conditions apply:
\begin{align}\label{rb2}
\epsilon_0 i\omega \underline{E}_\textrm{r}^{(\mathrm{s})}(R-d-\delta,\phi)&=\underline{j}_\textrm{r}(r,\phi)+\epsilon_0 i\omega \underline{E}^{(\mathrm{p})}_\textrm{r}(R-d-\delta,\phi)\\
\epsilon i\omega \underline{E}^{(\mathrm{d})}_{r}(R-d,\phi)&=\epsilon_0i\omega \underline{E}^{(\mathrm{s})}_{r}(R-d,\phi). 
\end{align}
The boundary condition for $\underline{E}_{\phi}^{(\mathrm{d})}$ on the resonators wall requires special attention. For $r=R$ the field $\underline{E}^{(\mathrm{d})}$ vanishes for all $\phi$ (idealized metal surface), except for $\lvert\phi\rvert< d_\textrm{s}/2R$ (gap area). According to chapter 2, an electrostatic regime is assumed within the gap and $\underline{E}^{(\mathrm{d})}$ can described by the field of a capacitor at that position. Therefore the boundary condition takes the following form, with gap size $d_\textrm{s}$ and the voltage $u$ at the gap:
\[
\underline{E}_{\phi}^{(\mathrm{d})}(R,\phi)=
\begin{cases}
     0 & \text{for } \lvert\phi\rvert\geq \arctan(d_\textrm{s}/2R)\approx d_\textrm{s}/2R \\
     u/\textrm{d}_s & \text{for } \lvert\phi\rvert< \arctan(d_\textrm{s}/2R)\approx d_\textrm{s}/2R.
   \end{cases}
	\]
This represents a periodic rectangle function, with period $2\pi$. It can be expanded into a Fourier series. By doing so, the boundary condition takes the following form:
\begin{align}\label{rb1}
\underline{E}_{\phi}(R,\phi)=\frac{u}{d_S 2\pi}+\sum_{m=1}^{\infty}\cos(m\phi)\frac{2uR\sin(\frac{m d_\textrm{S}}{2R})}{{d_\textrm{S}}^2 m \pi}.
\end{align}
Solving the equations \eqref{eq:221}-\eqref{eq:222} and taking the boundary conditions into account, it leads to the following fields for the plasma bulk region:
\newpage
\begin{align}
\underline{B}^{(\mathrm{p})}_z(r,\phi)&=\sum_{m=0}^{\infty}C_m J_m(r\kappa)\cos(m\phi) \label{eq:11} \\
\underline{E}^{(\mathrm{p})}_r(r,\phi)&=\frac{-c^2 m (-i\nu+\omega)}{r (\epsilon_r \nu\omega+i\epsilon_r \omega^2+i\omega_{pe}^2)}\sum_{m=0}^{\infty}C_m J_m(r\kappa)\sin(m\phi) \\
\underline{E}^{(\mathrm{p})}_{\phi}(r,\phi)&=\frac{c^2 (\nu+i\omega)}{\epsilon_r \omega(-i\nu+\omega)-\omega_{pe}^2}\frac{\kappa}{2}\sum_{m=0}^{\infty}C_m(J_{m-1}(r\kappa)-J_{m+1}(r\kappa))\cos(m\phi)\\
\underline{j}_r(r,\phi)&=\frac{\epsilon_0 \omega_{pe}^{2}}{\nu+i\omega}\underline{E}_r(r,\phi)\\
\underline{j}_\phi(r,\phi)&=\frac{\epsilon_0 \omega_{pe}^{2}}{\nu+i\omega}\underline{E}_\phi(r,\phi).\label{eq:21}
\end{align}
The relation between $\omega$ and $\kappa$ takes the following form:
\begin{align} 
\kappa^2=-\frac{\omega(\omega^2-i\nu\omega-\omega_{pe}^2)}{c^2(\omega-i\nu)}. \label{eq:kappa}
\end{align}
As the Fourier coefficients are part of $C_m$ (which can be found in the appendix) the expressions for $\underline{\vec{B}}^{(\mathrm{p})}(r,\phi)$, $\underline{\vec{E}}^{(\mathrm{p})}(r,\phi)$ and $\underline{\vec{j}}(r,\phi)$ consist of an infinite number of modes. The physical meaning of the individual modes is discussed later.

\subsection{Sheath modulation}
A constant sheath length of $\delta=5\times10^{-5}\,\textrm{m}$ is assumed. Based on a simple step model a typical length scale for the sheath modulation can be estimated by:   
\begin{align}
\Delta=\frac{j}{\textrm{e} n_e \omega_{\textrm{RF}}}.
\end{align}
Here $j$ represented the perpendicular component of the current that is flowing through the sheath. 
For the considered plasma density ($n_\textrm{e}\approx 10^{19} \textrm{m}^{-1}$) it is well justified to neglect $\Delta$ as $\frac{\Delta}{\delta}\ll10^{-3}$. Even for a considerably lower plasma density the sheath modulation is negligible.  

\subsection{Admittance and absorbed power}
In the next step the current $I$ that is being fed into the resonator is calculated. It is proportional to the discontinuity of the magnetic field strength $[\vec{H}\times\vec{n}]\propto I$ at the boundary between the resonator and the gap area. The current results from integrating the Maxwell-Ampère equation over an area $F=l\times2\epsilon_\textrm{F}$ with sides $l=l_2$ and $2\epsilon_\textrm{F}$. Here $F$ is a contour containing the edge between cylinder and the capacitor gap. As the surface normal of $F$ is aligned in $-\vec{e}_{r}$ direction, $I$ will completely flow through $F$ and it follows: 
\begin{align}
\int_{F}\nabla\times\vec{\underline{H}}\,\textrm{d}\vec{f}=\int_{F}\left( i\omega \vec{\underline{D}}+\vec{\underline{j}}_\textrm{f}\right)\,\textrm{d}\vec{f}=\underline{I}
\end{align}
Here $\vec{\underline{D}}$ represents the electric displacement field and $\vec{\underline{j}}_\textrm{f}$ represents the free current. Using Stoke's theorem this can be written as an integral over the boundary of $F$:
\begin{align}
\underline{I}=\oint_{\partial F}\vec{\underline{H}}\,\textrm{d}\vec{s}.
\end{align}
The integration path consists of four individual ways:
\begin{align} \label{eq:223}
\underline{I}=\int_{0}^{l}\underline{H}_\textrm{z}(R,\phi_\textrm{gap}+\epsilon_\textrm{F}/R)\,\textrm{d}z+\int_{\phi_\textrm{gap}-\epsilon_\textrm{F}/R}^{\phi_\textrm{gap} +\epsilon_\textrm{F}/R}\underline{H}_\textrm{z}(R,\phi)\,\textrm{d}\vec{\phi} \nonumber \\
 +\int_{l}^{0}\underline{H}_\textrm{z}(R,\phi_\textrm{gap}-\epsilon_\textrm{F}/R)\,\textrm{d}z+\int_{\phi_\textrm{gap}+\epsilon_\textrm{F}/R}^{\phi_\textrm{gap}-\epsilon_\textrm{F}/R}\underline{H}_\textrm{z}(R,\phi)\,\textrm{d}\vec{\phi}.
\end{align}
In cylindrical coordinates $(R,\phi_\textrm{gap})$ represents the position of the edge where the inner surface of the cylindrical tube an the gap capacitor meet. Due to $\vec{\underline{H}}\bot\vec{e}_{\phi}$ only the first and third term of \eqref{eq:223} contribute to the result:
\begin{align}
\underline{I}=\underline{H}_\textrm{z}(R,\phi_\textrm{gap}+\epsilon_\textrm{F})-\underline{H}_\textrm{z}(R,\phi_\textrm{gap}-\epsilon_\textrm{F}).
\end{align}
The jump of the magnetic field strength is equal to $\vec{\underline{B}}$ at the capavitor edge. Thus: 
\begin{align}
\underline{H}_\textrm{z}(R,\phi_\textrm{gap}+\epsilon_\textrm{F})-\underline{H}_\textrm{z}(R,\phi_\textrm{gap}-\epsilon_\textrm{F})=\frac{1}{\mu_0}\underline{B}_\textrm{z}(R,\phi_\textrm{gap}).
\end{align}
Finally an expression for the plasma admittance $Y_\textrm{P}$ can be gained:
\begin{align}
Y_\textrm{P}=\frac{I}{u}=\frac{1}{u}\frac{l}{\mu_0}\underline{B}_\textrm{z}(R,\phi_\textrm{gap}).
\end{align}
The admittance is expressed as a function of $n_{\textrm{e}}$ and $\omega$, as $\kappa$ is a function of $\omega$ and $\omega_{\textrm{pe}}$ (apparent from equation (\ref{eq:kappa})) and $\omega_{\textrm{pe}}=\sqrt{n_\textrm{e}e^2 / \epsilon_0 m_\textrm{e}}$. In addition $Y_\textrm{P}$ consists of an infinite number of modes as it includes the expression of $\vec{B}$:
\begin{align}\label{eq:padm}
Y_\textrm{P}(\omega,n_e)=\sum_{m=0}^{\infty}Y_{m}(\omega,n_{\textrm{e}}).
\end{align} 
The admittance of the plasma from equation (\ref{eq:padm}) is used to evaluate eq. (\ref{eq:impedance}). The resulting expression of $Y_\textrm{S}(\omega,n_{\textrm{e}})$ contains fundamental information of the combined system of plasma and resonator and its frequency dependent behavior. It can be related to the lumped element model presented in Fig 2. The individual modes of the plasma admittance $Y_{\textrm{P},m}(\omega,n_{\textrm{e}})$ can be identified with different branches in the parallel circuit on the right in Fig 2 and different discharge regimes. This is discussed in more detail later. The real part of $Y_\textrm{S}(\omega,n_{\textrm{e}})$ is shown in Fig \ref{fig:antragAdmittance}. For a given microwave source with amplitude $u_{\textrm{S}}$ the power dissipated in the plasma finally takes the form:
\begin{align}
P_{\textrm{abs}}(\omega,n_{\textrm{e}})&=\frac{1}{2}|u_{\textrm{S}}|^2 \operatorname{Re}(Y_\textrm{S}(\omega,n_{\textrm{e}}))\\
\end{align}

\pagebreak

\pagebreak

\section{Plasma model}

In this section global balance equations is used to describe certain plasma parameters. The term "global" refers to a volume averaged description. This kind of model is commonly employed to predict spatially averaged parameters if the spatial information is not of interest \cite{ref13}. The global model is based on a particle balance and a power balance equation. At first the particle balance is considered. In the steady state regime the number of electrons $N_{\textrm{e}}$ is constant. Thus, the number of electrons gained is in balance with the number of electrons lost:
\begin{align} 
\frac{\textrm{d}N_{\textrm{e}}}{\textrm{d}t}=K_{\textrm{iz}}(T_{\textrm{e}})N_{\textrm{e}}-\frac{1}{\tau_{\textrm{loss}}}N_{\textrm{e}}=0\label{eq:224}. 
\end{align}
Here $K_{\textrm{iz}}$ represents the rate coefficient for ionization as a function of the electron-temperature $T_\textrm{e}$, and $\tau_{\textrm{loss}}$ is an effective time constant that describes the decay processes. Depending on the plasma parameters various ionization processes play a role (direct ionization, ionization through various exited states). The corresponding rate coefficients are to be selected accordingly. As the number of electrons is time independent, equation (\ref{eq:224}) can be used for the determination of $T_\textrm{e}$. Secondly, the power balance is considered, with the absorbed power, $P_{\textrm{\textrm{abs}}}$, being equal to the power lost:
\begin{align}
\frac{\textrm{d}}{\textrm{d}t} \left(\frac{3}{2} n_{\textrm{e}} T_{\textrm{e}}\right) =P_{\textrm{abs}}-K^{*}(T_{\textrm{e}})\epsilon^{*}N_{\textrm{e}}=0. \label{eq:101}
\end{align}
Where generally $K^{*}(T_{\textrm{e}})$ is a rate coefficient and $\epsilon^{*}$ the corresponding energy. Following the considerations above, a power loss model for an argon plasma is being developed. The total power loss $P_{\textrm{loss}}(n_e)$ is finally a function of $n_e$. In order to determine $P_\textrm{loss}$ more precisely, the flux of particles with respect to the boundary surfaces is considered in cylindrical coordinates. For radial losses the surface of the cylinder wall  $A_\textrm{M}=2 \pi R l$ represents the boundary surface. The question arises if the axial losses can be treated in a similar way. Since the considered plasma operates as a jet, there are no axial walls and plasma processes are not strongly limited to the extension of the resonator block. Therefore, it is expected that the radial losses are dominant in comparison to the axial losses. We neglect the axial losses and assume a constant density profile along the z-axis as the plasma jet extends much beyond the resonator length. Under this assumption the considered wall flux of particles $\Psi_w$ takes the form:
\begin{align}
\Psi_{\textrm{w}}=v_\textrm{B} n_0 A_\textrm{M}h.
\end{align}
Here $v_\textrm{B}$ denotes the Bohm velocity. Further on $n_0$ denotes the electron density short of the sheath. The so called $h$-factor is introduced in order to express the wall flux $\Psi_w$ as a function of the average electron density $n_{e}$ \cite{ref14}. For this simplified model, the considered geometry and a reference pressure of $100\,\textrm{Pa}$ the following $h$-factor is used \cite{ref10}.
\begin{align}
h=0.80 \left(4+\frac{R}{\lambda_\textrm{i}}\right)^{-1/2}.
\end{align}
Here $\lambda_\textrm{i}$ denotes the mean free path for ions. For further calculations the collisional energy loss per electron-ion pair created, $\varepsilon_{\textrm{c}}$, is considered. It is defined by \cite{ref10}:
\begin{align}
K_{\textrm{c}}\varepsilon_{\textrm{c}}=K_{\textrm{iz}}\varepsilon_{\textrm{iz}}+K_{\textrm{ex}}\varepsilon_{\textrm{ex}}+K_{\textrm{el}}\frac{3 m}{M} T_\textrm{e}.
\end{align}
In this equation the loss of electron energy due to ionization, $\varepsilon_{\textrm{iz}}$, excitation, $\varepsilon_{\textrm{ex}}$, and the mean energy for elastic scattering (with neutrals), $\frac{3 m}{M} T_\textrm{e}$, is taken into account. Stepwise ionization is neglected as the quantitative difference to direct ionization is less than 20\%, according to the model of Kemaneci et al., \cite{ref15}. Here $K_{\textrm{iz}}$, $K_{\textrm{ex}}$ and $K_{\textrm{el}}$ represent the corresponding rate coefficients, while $m$ is the electron mass and $M$ the neutral particle mass. The values of these rate coefficients are $K_{\textrm{iz}}=2.34\times10^{-14}\textrm{T}_\textrm{e}^{0.59}\textrm{e}^{-17.44/\textrm{T}_\textrm{e}}\,\textrm{m}^3/s$, $K_{\textrm{ex}}=2.48\times10^{-14}\textrm{T}_\textrm{e}^{0.33}\textrm{e}^{-12.78/\textrm{T}_\textrm{e}}\,\textrm{m}^3/s$ and $K_{\textrm{el}}=2.336\times10^{-14}\textrm{T}_\textrm{e}^{1.609}\textrm{e}^{0.0618\ln(\textrm{T}_\textrm{e})^2-0.1171\ln(\textrm{T}_\textrm{e})^3}\,\textrm{m}^3/s$, according to [15]-[16]. For argon $\varepsilon_{\textrm{iz}}=15.76\,\textrm{eV}$ and $\varepsilon_{\textrm{ex}}=12.14\,\textrm{eV}$ [10]. In addition, the energy loss for ions, $e V_\textrm{sh}$, and the average kinetic energy carried out by lost electrons, $2 T_{\textrm{e}}$, is to be accounted. The sheath-voltage $V_\textrm{sh}$ represents the potential difference caused by the unequal charge distribution in the sheath. The elementary charge is denoted with $e$. In summary the total energy loss $E_\textrm{loss}$ per electron-ion pair created is given by:
\begin{align}\label{xy1}
E_\textrm{loss}=\varepsilon_{\textrm{iz}}+\frac{K_{\textrm{ex}}}{K_{\textrm{iz}}}\varepsilon_{\textrm{ex}}+\frac{K_{\textrm{el}}}{K_{\textrm{iz}}}\frac{3 m}{M} T_\textrm{e}+e V_\textrm{sh}+2T_{\textrm{e}}.
\end{align} 
Multiplication of eq. (\ref{xy1}) with the flux of particles $\psi_\textrm{w}$ leads to an expression for the total loss power:
\begin{align}
P_{\textrm{loss}}(n_{\textrm{e}})=n_{\textrm{e}} v_{\textrm{B}}A_{\textrm{M}}h \left(\varepsilon_{\textrm{iz}}+\frac{K_{\textrm{ex}}}{K_{\textrm{iz}}}\varepsilon_{\textrm{ex}}+\frac{K_{\textrm{el}}} {K_{\textrm{iz}}}\frac{3 m}{M} T_\textrm{e}+e V_\textrm{sh}+2 T_{\textrm{e}}\right).
\end{align}
Using the presented coefficients for argon, equations \eqref{eq:224} gives an electron temperature of $T_{\textrm{e}}=2.5\,\textrm{eV}$.

\section{Characteristics of the system}  
\begin{figure}[htbp]
\centering
\iffigures		
   	\includegraphics[width=0.89\textwidth]{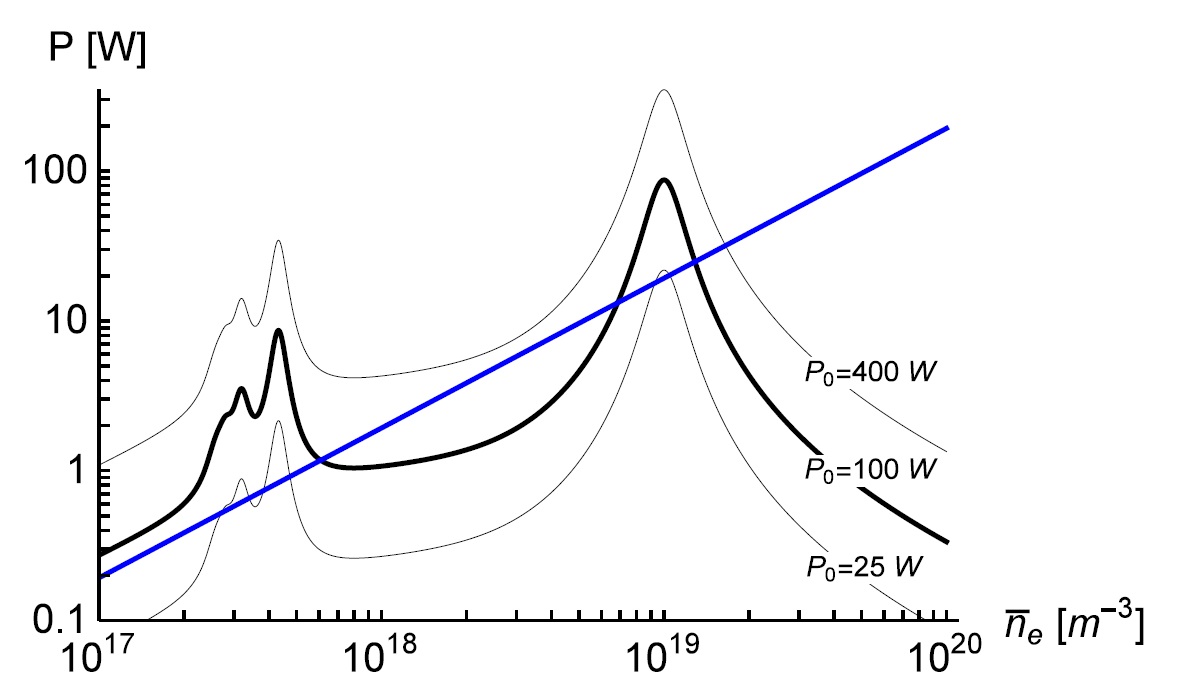}
\else 
		
\fi 
\caption{Absorbed power $P_{\textrm{abs}}$ for a "`single jet"' as a function of averaged $n_e$ for various source-powers $P_0$, as well as loss power $P_{\textrm{loss}}$(blue line). Intersection of both curves represent stationary operating points.}
	\label{fig:pow1}
\end{figure}

\begin{figure}
\centering
	  \iffigures		
	     	\includegraphics[width=0.99\textwidth]{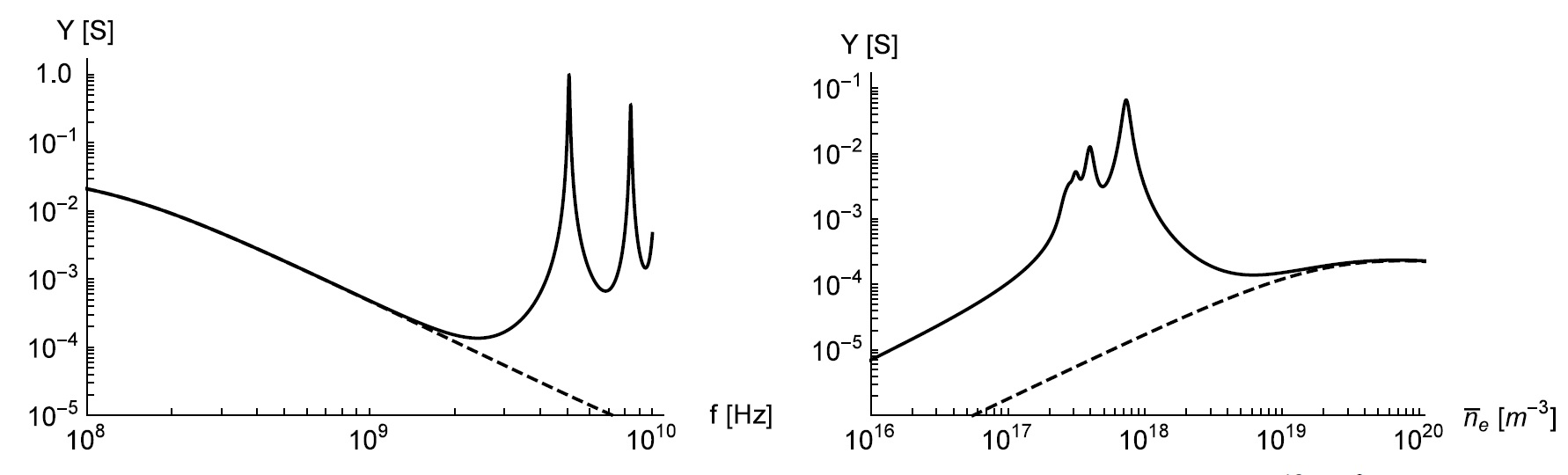}
	\else 
		
	\fi 
	\caption{Real part of the complex admittance $Y(\omega,n_{\textrm{e}})$ plotted as a function of frequency $f$ for a constant $n_e=10^{19}\textrm{m}^{-3}$ (left) and as a function of $n_e$ for a constant frequency $\omega=2 \pi\times2.45\,\textrm{GHz}$(right). The dashed line represents the fundamental $Y_0$ mode, while the solid line represents the sum of the modes.}
	\label{fig:antragAdmittance}
\end{figure}

\begin{figure}[htbp] 
	\centering
  \iffigures		
	     	\includegraphics[width=0.99\textwidth]{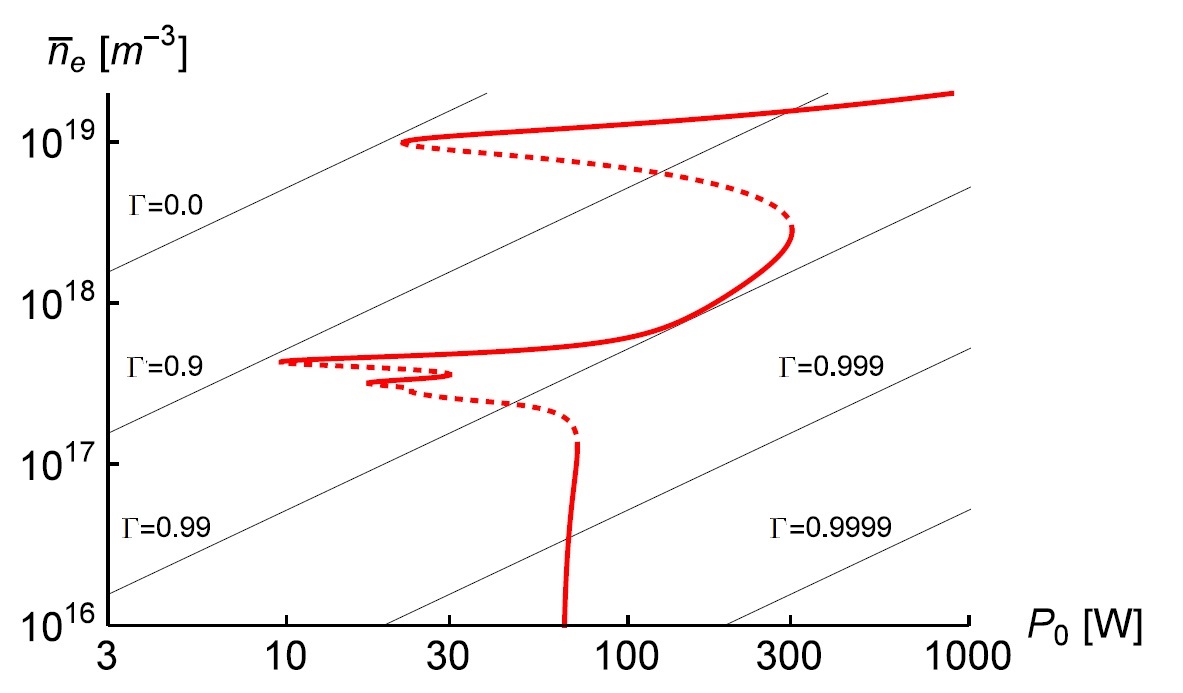}
	\else 
		
	\fi 
	\caption{Stable (solid) and unstable (dashed) operating points with corresponding $n_{\textrm{e}}$ and reflection coefficient $\left|\Gamma\right|$.}
	\label{fig:antrag5}
\end{figure}

\begin{figure}[htbp]
	\centering
\iffigures		
   		\includegraphics[width=0.99\textwidth]{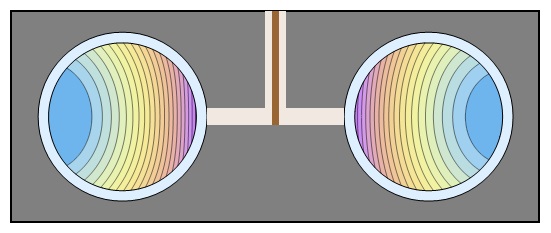}
\else 
		
\fi 
\caption{Electric field lines for $n_e\approx 10^{17}$.\\
	The corresponding mode is called E-mode.}
	\label{fig:bild55}
\end{figure}

\begin{figure}[htbp]
	\centering
  \iffigures		
	     \includegraphics[width=0.99\textwidth]{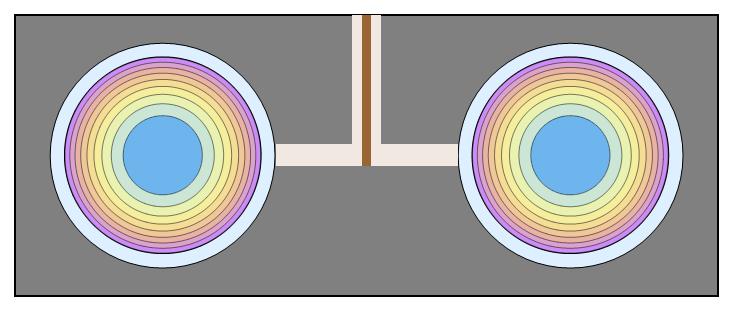}
	\else 
		
	\fi 
	\caption{Electric field lines for $n_e\approx 10^{19}$.\\
	The corresponding mode is called H-mode. In this case the power \\ is mainly inductively coupled to the plasma.}
	\label{fig:bild57}
\end{figure}

In this section fundamental insights into the characteristics of the system will be gained. To investigate the power balance $P_{\textrm{loss}}$ and $P_{\textrm{abs}}$ are plotted as functions of $n_\textrm{e}$. The resulting plot is shown in Fig \ref{fig:pow1}. Intersections of both curves represent operating points and meet the condition:
\begin{align}\label{eq:5}
P_{\textrm{\textrm{abs}}}(n_{\textrm{e}})=P_{\textrm{loss}}(n_{\textrm{e}}). 
\end{align}
It is assumed that an operating point can be characterized as stable if the following criterion is met:
\begin{align}\label{eq:444}
\frac{\partial P_{\textrm{abs}}}{\partial n_{e}}<\frac{\partial P_{\textrm{loss}}}{\partial n_{e}}.
\end{align}
According to Fig \ref{fig:pow1} the absorbed power $P_{\textrm{abs}}$ shows a nonlinear behavior as a function of $n_{\textrm{e}}$ resulting in multiple solutions of equation \ref{eq:5}. Thus, a hysteresis is expected to occur with the variation of the plasma parameters and input power. The set
\begin{align}\label{eq6}
\left\{(n_{\textrm{e}},P)\,\lvert\,P:=P_{\textrm{abs}}(n_{\textrm{e}})\overset{!}{=}P_{\textrm{loss}}(n_{\textrm{e}})\right\} 
\end{align}
represents all possible combinations of $n_{\textrm{e}}$ and $P$ for operation in power balance. In Fig \ref{fig:antrag5} the result is shown graphically. Stable and unstable operating points are marked differently. In addition, the real value of the reflection coefficient $\lvert \Gamma\rvert$ is plotted in figure \ref{fig:antrag5}, with:
\begin{align}
\Gamma=\frac{Z_\textrm{S}-Z_0}{Z_\textrm{S}+Z_0}.
\end{align}
The impedance of the combined system of plasma and resonator is given by $Z_\textrm{S}=Y_\textrm{S}(\omega,n_{\textrm{e}})^{-1}$. Finally the pattern of the field lines is plotted for two characteristic values of the electron density ($n_\textrm{e} \approx 10^{17} \textrm{m}^{-3}$ and $n_\textrm{e}\approx10^{19} \textrm{m}^{-3}$). The resulting plots are shown in Figs \ref{fig:bild55} and \ref{fig:bild57}.
\newline
\newline
In conclusion the following insights can be gained: 
\begin{itemize}
\item Two clearly separated operating regimes can be identified. One regime is located at a value of $n_\textrm{e} \approx 10^{17} \textrm{m}^{-3}$ and the second regime is located at a value of $n_\textrm{e}\approx10^{19} \textrm{m}^{-3}$.
\item The mode $Y_{0}(\omega,n_{\textrm{e}})$ constitutes the dominant contribution in the second (high $n_{\textrm{e}}$) regime, while $Y_{1}(\omega,n_{\textrm{e}})$ and higher modes dominate the first (low $n_{\textrm{e}}$) regime (as resulted from Fig 4).
\item For the low $n_{\textrm{e}}$ case the field line patterns are similar to a field of a capacitor and it can be assumed the energy is mainly capacitively coupled to the system. For the high $n_{\textrm{e}}$ case circular currents are present and a mainly inductively coupled energy transfer can be assumed.
\item The low $n_{\textrm{e}}$ mode requires a minimum of $P_0=10\,\textrm{W}$ and appears to be relatively poor in terms of efficiency $(\left|\Gamma\right|\approx0.9)$. The high $n_{\textrm{e}}$ mode requires a minimum of $P_0=25\,\textrm{W}$ and gives a significantly higher efficiency $(\left|\Gamma\right|\approx0.1)$.
\item Hysteresis occurs when cycling between the two discharge regimes. 
\end{itemize}
Consequently the high $n_{\textrm{e}}$ mode is called inductive mode or H-mode and corresponds to the inductive branch ($m=0$) in the lumped element model in Fig 2. The low $n_{\textrm{e}}$ mode is called capacitive mode or E-mode and corresponds to the capacitive branch ($m=1$) in the lumped element model in Fig 2. Modes of higher order provide a far less important contribution. Particularly, during H-mode operation they can be neglected. 
\newline
\newline
First experimental results have been gathered with the MMWICP \cite{ref11}-\cite{ref12}. The predicted behavior is in good agreement with the experimental results concerning fundamental aspects of the system. This regards particularly the existence of multiple modes, different discharge regimes and the values of $n_{\textrm{e}}$ as well as the predicted 2D pattern of field lines (the H-mode is optically proven for the MMICP and a nitrogen discharge [11]). A quantitative comparison to experimental results will be subject to further research when a global model for nitrogen and additional diagnostics are available.

\pagebreak

\section{Summary and conclusion}

In this paper a Miniature Microwave ICP plasma jet (MMWICP) is theoretically investigated. An electromagnetic model for the absorbed power is delivered based on fundamental field equations. This leads to a Fourier series representation of the electromagnetic fields with an infinite number of modes ordered by the azimuthal wave number m. In addition a global model for the power loss is derived. By equating both calculated powers, $P_{\textrm{abs}}=P_{\textrm{loss}}$, it is possible to identify stable operating points. With respect to the electron density, two clearly separated discharge regimes are identified. The corresponding field lines patterns are plotted. The results are used to determine the physical meaning of the discharge regimes. The high electron density regime requires $n_{\textrm{e}}\approx10^{19}\,\textrm{m}^{-3}$ and represents inductive coupling. It is identified as H-mode. In this case strong circular currents are present and the mode $m=0$ is dominating. The low electron density regime requires $n_{\textrm{e}}\approx5\cdot10^{17}\,\textrm{m}^{-3}$ and represents capacitive coupling. Consequently it is identified as E-mode. In this case the field line pattern are similar to a capacitor and the mode $m=1$ is dominating. Due to the nonlinear behavior of $P_{\textrm{abs}}$, the equation $P_{\textrm{abs}}=P_{\textrm{loss}}$ has multiple solutions. Therefore hysteresis effects appear at different points, when cycling between E- and H-mode. The efficiency is also depending on the discharge regimes. During E-mode operation the efficiency ist low: $(\left|\Gamma\right|\approx0.9)$. Only a low degree of the power is transported into the plasma. In contrast to that operation in E-mode provides high efficiency and low reflection: $(\left|\Gamma\right|\approx0.09)$. A comparison to existing experiments is limited to the electromagnetic aspects as experimental results are only available for nitrogen. A quantitative comparison to experimental results will be subject to further research when a global model for nitrogen and additional diagnostics are available. In conclusion the feasibility of inductive power coupling for the MMWICP resonator is shown and its electromagnetic characteristics are examined and presented. 

\section{Acknowledgements}

The authors gratefully acknowledge support by Deutsche Forschungsgemeinschaft DFG via 
the project 389090373. 

\pagebreak

\section*{Appendix}
The constant $C_m$ results from solving equations \eqref{eq:221}-\eqref{eq:222}. It contains the Fourier coefficients and takes the following form:    
\small
\begin{multline*}
C_m=4 c^4 R \omega  \epsilon_{\textrm{r}} \left(i \nu  \omega -\omega ^2+\omega_{\textrm{pe}}^2\right) Y_m\left(C_1\right) Y_m\left(C_2\right) \\ 
\times\left( \begin{cases}
     \frac{u}{d_S 2\pi} & \text{for } m=0 \\
     \frac{2uR\sin(\frac{m d_\textrm{S}}{2R})}{{d_\textrm{S}}^2 m \pi} & \text{for } m\geq1
   \end{cases} \right)\left(\pi ^2\left(i c \omega  (d+\delta -R) \left(i \nu  \omega -\omega ^2+\omega_{\textrm{pe}}{}^2\right) J_m(-(d-R+\delta ) \kappa )\right.\right. \\
\times J_{m-1}\left(C_1\right) Y_m\left(C_1\right)-J_m\left(C_1\right) Y_{m-1}\left(C_1\right) J_m\left(C_2\right) Y_{m-1}\left(C_2\right)-J_{m-1}\left(C_2\right) Y_m\left(C_2\right) \\
-\left(c \left(R \omega  \sqrt{\epsilon_{\textrm{r}}} Y_{m-1}\left(C_3\right)-c m Y_m\left(C_3\right)\right) J_m\left(C_2\right)\right. +\left.c \left(c m J_m\left(C_3\right)-R \omega  \sqrt{\epsilon_{\textrm{r}}} J_{m-1}\left(C_3\right)\right) Y_m\left(C_2\right)\right) \\
\times \left(c \omega  \sqrt{\epsilon_{\textrm{r}}} (R-d) Y_m\left(C_4\right) Y_{m-1}\left(C_2\right)\right. +\left.\left.c \left(c m (\epsilon_{\textrm{r}}-1) Y_m\left(C_4\right)+\omega  \epsilon_{\textrm{r}} (d-R) Y_{m-1}\left(C_4\right)\right) Y_m\left(C_2\right)\right)\right) \\
+c \left(c \omega  (\nu +i \omega ) (\kappa  (d+\delta -R) J_{m-1}(-(d-R+\delta ) \kappa )+m J_m(-(d-R+\delta ) \kappa )) Y_m\left(C_1\right)\right. \\
+i \left(i \nu  \omega -\omega ^2+\omega_{\textrm{pe}}{}^2\right) J_m(-(d-R+\delta ) \kappa ) [\omega  (d+\delta -R) Y_{m-1}\left(C_1\right)+c m Y_m\left(C_1\right)]\left(J_m\left(C_5\right)\left(c^2 \omega  \sqrt{\epsilon_{\textrm{r}}} (d-R) Y_m\left(C_4\right)\right)\right) \\
\times\left(c m Y_m\left(C_3\right)-R \omega  \sqrt{\epsilon_{\textrm{r}}} Y_{m-1}\left(C_3\right)\right) J_m\left(C_2\right) Y_{m-1}\left(C_2\right)-J_{m-1}\left(C_2\right) Y_m\left(C_2\right) \\
-\left(c \left(R \omega  \sqrt{\epsilon_{\textrm{r}}} Y_{m-1}\left(C_3\right)-c m Y_m\left(C_3\right)\right) J_m\left(C_2\right)\right.+\left.c \left(c m J_m\left(C_3\right)-R \omega  \sqrt{\epsilon_{\textrm{r}}} J_{m-1}\left(C_3\right)\right) Y_m\left(C_2\right)\right)\\
\times\left(c \omega  \sqrt{\epsilon_{\textrm{r}}} (R-d) Y_m\left(C_4\right) Y_{m-1}\left(C_2\right)\right.+\left.\left.c \left(c m (\epsilon_{\textrm{r}}-1) Y_m\left(C_4\right)+\omega  \epsilon_{\textrm{r}} (d-R) Y_{m-1}\left(C_4\right)\right) Y_m\left(C_2\right)\right)\right)\\
\times Y_m\left(C_5\right)\left(-c^2 \omega  \sqrt{\epsilon_{\textrm{r}}} (d-R) J_m\left(C_4\right) \left(c m Y_m\left(C_3\right)-R \omega  \sqrt{\epsilon_{\textrm{r}}} Y_{m-1}\left(C_3\right)\right)\right. J_m\left(C_2\right) Y_{m-1}\left(C_2\right)-J_{m-1}\left(C_2\right) Y_m\left(C_2\right) \\
+ \left(c \left(R \omega  \sqrt{\epsilon_{\textrm{r}}} Y_{m-1}\left(C_3\right)-c m Y_m\left(C_3\right)\right) J_m\left(C_2\right)\right.+ \left.c \left(c m J_m\left(C_3\right)-R \omega  \sqrt{\epsilon_{\textrm{r}}} J_{m-1}\left(C_3\right)\right) Y_m\left(C_2\right)\right) \\ 
\end{multline*}
The meaning of the following abbreviations applies:
\begin{align*}
C_1&=-\frac{(d-R+\delta ) \omega }{c} \\
C_2&=\frac{(R-d) \sqrt{\epsilon_{\textrm{r}}} \omega }{c} \\
C_3&=\frac{R \sqrt{\epsilon_{\textrm{r}}} \omega }{c} \\
C_4&=\frac{(R-d) \omega }{c} \\
C_5&=-\frac{\omega  (d+\delta -R)}{c}. \\
\end{align*}

\pagebreak

\pagebreak

\section{Figures}

\pagebreak

\pagebreak

\pagebreak

\pagebreak

\pagebreak

\pagebreak

\pagebreak

\end{document}